\documentclass[final,twocolumn,prl,aps,floats,amsfonts,amssymb,showpacs,superscriptaddress]{revtex4}
\usepackage[utf8]{inputenc}
\usepackage{graphicx}
\usepackage{color}
\usepackage{psfrag}
\usepackage{gensymb}
\usepackage{enumerate}
\usepackage{epstopdf}
\usepackage{amssymb}
\usepackage{amsmath}
\usepackage{wasysym}
\usepackage{hyperref}
\usepackage{csquotes}
\usepackage{color}
\usepackage{enumerate}

\begin{document}
\title{  Maximum Kolmogorov-Sinai entropy vs minimum mixing time in Markov chains}

\author{M. Mihelich} 
\email{martin.mihelich@cea.fr}
\affiliation{
Laboratoire SPHYNX, Service de Physique de l'Etat Condens\'e, DSM,
CEA Saclay, CNRS UMR 3680, 91191 Gif-sur-Yvette, France
}

\author{B. Dubrulle} 
\affiliation{
Laboratoire SPHYNX, Service de Physique de l'Etat Condens\'e, DSM,
CEA Saclay, CNRS UMR 3680, 91191 Gif-sur-Yvette, France
}

\author{D. Paillard} 
\affiliation{Laboratoire des Sciences du Climat et de l'Environnement, IPSL, CEA-CNRS-UVSQ, UMR 8212, 91191 Gif-sur-Yvette, France}

\author{Q. Kral} 
\affiliation{Institute of Astronomy, University of Cambridge, Madingley Road, Cambridge CB3 0HA, UK}
\affiliation{LESIA-Observatoire de Paris, UPMC Univ. Paris 06, Univ. Paris-Diderot, 92195 Meudon Cedex, France}

\author{D. Faranda}
\affiliation{Laboratoire des Sciences du Climat et de l'Environnement, IPSL, CEA-CNRS-UVSQ, UMR 8212, 91191 Gif-sur-Yvette, France}

\date[Writing in progress: ]{\today}


\begin{abstract}
We establish a link between the maximization of Kolmogorov Sinai entropy (KSE) and the minimization
of the mixing time for general Markov chains. Since the maximisation of KSE is analytical and easier to compute in general than mixing time, this link provides a new faster method 
to approximate the minimum mixing time dynamics. It could be interesting
in computer sciences and statistical physics, for computations that use random walks on graphs that can be represented as Markov chains.
\end{abstract}
\maketitle

Many modern techniques of  physics, such as computation of path integrals, now rely on random walks on graphs that can be represented as Markov chains.
Techniques to estimate the number of steps in the chain
to reach the stationary distribution (the so-called \enquote{mixing time}), are of great
importance in obtaining estimates of running times of such sampling algorithms \cite{bhakta2013mixing} (for a review of existing techniques, see e.g. \cite{guruswami2000rapidly}). On the other hand, studies of the link between the topology of the graph and the diffusion properties of the random walk on this graph are often based on the entropy rate, computed using the Kolmogorov-Sinai entropy (KSE)\cite{gomez2008entropy}. For example, one can  investigate dynamics on a network maximizing the KSE to study  optimal diffusion
\cite{gomez2008entropy}, or  obtain an algorithm to produce equiprobable paths on non-regular graphs \cite{burda2009localization}.\

In this letter, we establish a link between these two notions by showing that for a
system that can be represented by Markov chains, \textbf{a non trivial
relation exists between the maximization of KSE and the minimization
of the mixing time}. Since KSE are easier to compute in general than mixing time, this link provides a new faster method 
to approximate the minimum mixing time that could be interesting
in computer sciences and statistical physics and gives a physical meaning to the KSE. We first show that on average, the greater the KSE, the smaller the mixing time, and we correlated this result to its link with the transition matrix eigenvalues. Then, we show that the dynamics that maximises KSE is close to the one minimizing the mixing time, both in the sense of  the optimal diffusion coefficient and the transition matrix.\

Consider a network with $m$ nodes, on which a particle jumps randomly. This process can be described by a finite Markov chain defined by its adjacency matrix $A$ and its transition matrix $P$. $A(i,j)=1$ if and only if there is a link between the nodes $i$ and $j$ and 0 otherwise. $P=(p_{ij})$ where $p_{ij}$ is the probability for a particle in $i$ to hop on the $j$ node. Let us introduce the probability density at time $n$ $\mu_n=(\mu_n^i)_{i=1...m}$ where $\mu_n^i$ is the probability that a particle is at node $i$ at time $n$. Starting with a probability density $ \mu_0$, the evolution of the probability density writes:  $\mu_{n+1}=P^t\mu_{n}$ where $P^t$ is the transpose matrix of $P$.\\
 Within this paper, we assume that the Markov chain is irreducible and thus has a unique stationary state.

Let us define: 
\begin{equation}
\label{eqdn}
 d(n)= max{ || (P^t)^n\mu - \mu_{stat}|| \text{  } \forall \text{ } \mu }, 
\end{equation}
where $||.||$ is a norm on $\mathbb{R}^n$. For $ \epsilon >  0$, the mixing time, which corresponds to the time such that the system is within a distance $\epsilon$ from its stationary state  is defined as follows: 

\begin{equation}
\label{eq:mix1}
t(\epsilon)= \min{  n,  \, d(n) \leq \epsilon}.
\end{equation}

For a Markov chain the KSE takes the analytical form \cite{billingsley1965ergodic}:

\begin{equation}
\label{eqhks}
 h_{KS}=-\sum_{ij} \mu_{stat_{i}}p_{ij}\log(p_{ij}).
\end{equation}

\begin{figure}[hb!]
   \centering
   \includegraphics[width=10cm]{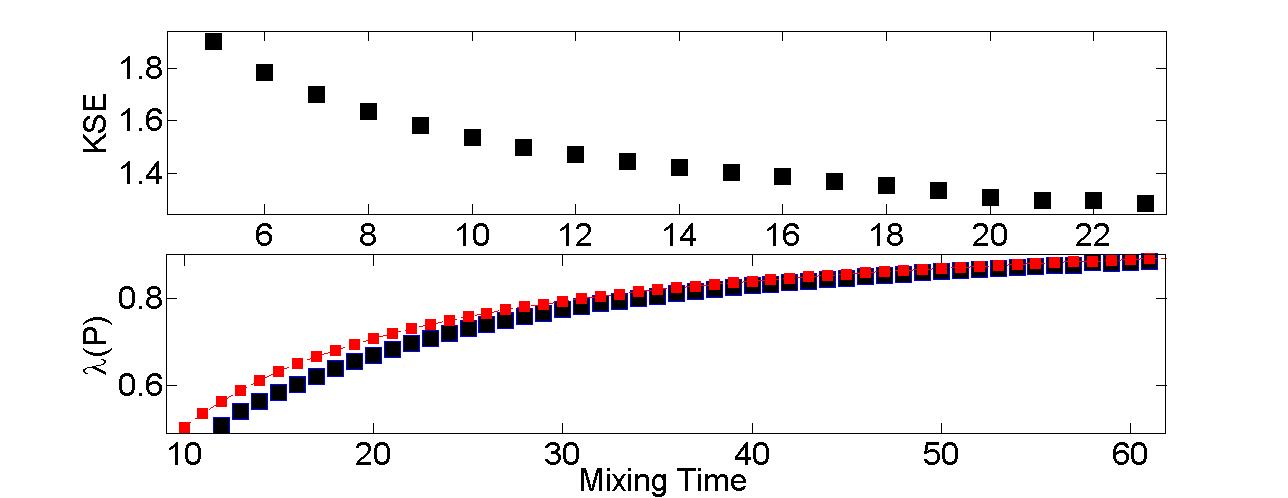}
  \caption{Averaged KSE versus mixing time (top) for $10^6$ random $m=10$ size matrices and averaged $\lambda(P)$ versus mixing time (bottom) for $10^6$ random $m=10$ size matrices in curve blue and $f(t)=\epsilon^{1/t}$ in red. $\epsilon=10^{-3}$ and the norm is chosen to be the euclidian one.}
\label{fig:KS1}
 \end{figure}

Random $m$ size Markov matrices are generated by assigning to each $p_{ij}$  ($i\neq j$) a random number between $0$ and $ \frac{1}{m}$ and $p_{ii}= 1-\sum_{j\neq i} p_{ij}$. The mean KSE is plotted versus the mixing time (Fig.~\ref{fig:KS1}) by working out $h_{KS}$ and $t(\epsilon)$ for each random matrix. (Fig.~\ref{fig:KS1}) shows that KSE is  on average a decreasing function of the mixing time.

We stress the fact that this relation is only true on average. We can indeed find two special Markov chains $P1$ and $P2$ such that $h_{KS}(P1) \leq h_{KS}(P2)$ and $t_1(\epsilon) \leq t_2(\epsilon)$. We illustrate this point further. \

The link between the mixing time and the KSE can be understood via their dependence as a function of  the transition matrix eigenvalues. A general irreducible transition matrix $P$ is not necessarily diagonalizable on $\mathbb{R}$. However, since $P$ is chosen randomly, it is almost everywhere diagonalizable on $\mathbb{C}$. According to Perron Frobenius theorem, the largest eigenvalue is 1 and the associated eigen-space is one-dimensional and  equal to the vectorial space generated by $\mu_\text{stat}$. Without loss of generality, we can label the eigenvalues in decreasing order of their module:

 $$1=\lambda_1 > \lvert \lambda_2 \rvert \geq....\geq \lvert \lambda_m \rvert \geq 0$$

 The  convergence speed toward $\mu_\text{stat}$ is given by the second maximum module of the eigenvalues of  $P$ \cite{boyd2004fastest}, \cite{pierre1999markov}:

$$ \lambda(P)=\max_{i=2...m}{ |\lambda_i|}= \lvert \lambda_2 \rvert$$

The eigenvalues $\lambda_1=1,...,\lambda_m$ of $P$ and $P^t$ being equal, let us denote their associated eigenvectors $\mu_1=\mu_\text{stat},...,\mu_m$. For any initial probability density $\mu_0$, we find:

\begin{equation}
\label{eqmu0}
 || (P^t)^n\mu_0 - \mu_\text{stat}|| \propto (\lambda(P))^n.
\end{equation}

According to Eqs.~(\ref{eqdn}) and (\ref{eq:mix1}), $\lambda(P)^{t(\epsilon)} \propto \epsilon$, i.e.  $\lambda(P) \propto \epsilon^{1/t(\epsilon)}$. Hence, the smaller $\lambda(P)$ the shorter the mixing time (Fig.~\ref{fig:KS1}). $h_{KS}$ being a decreasing function of $t(\epsilon)$ and $\lambda(P)$ being an increasing function of $t(\epsilon)$, we deduce that $h_{KS}$ is a decreasing function of $\lambda(P)$.\

This link between maximum KSE and minimum mixing time actually also extends naturally to optimal diffusion coefficients. Such a notion has been introduced by Gomez-Gardenes and Latora \cite{gomez2008entropy} in networks represented by a Markov chain depending on a diffusion coefficient. Based on the observation that in such networks,  KSE  has a maximum as a function of the diffusion coefficient, they define an optimal diffusion coefficient as the value of the diffusion corresponding to this maximum. In the same spirit, one could compute an optimal diffusion coefficient with respect to the mixing time, corresponding to the value of the diffusion coefficient which minimizes the mixing time -or equivalently the smallest second largest eigenvalue $\lambda(P)$. This would roughly correspond to the diffusion model reaching the stationary time in the fastest time.  To define such an optimal diffusion coefficient, we follow Gomez and Latora and vary the transition probability depending on the degree of the graph nodes. More accurately, if $k_i=\sum_j A(i,j)$ denotes the degree of node $i$, we set:

\begin{equation}
\label{eq:diff1}
 p_{ij}=\frac{A_{ij}k_j^\alpha}{\sum_j A_{ij}k_j^\alpha}.
\end{equation}

\begin{figure}[hb!]
   \centering
   \includegraphics[width=10cm]{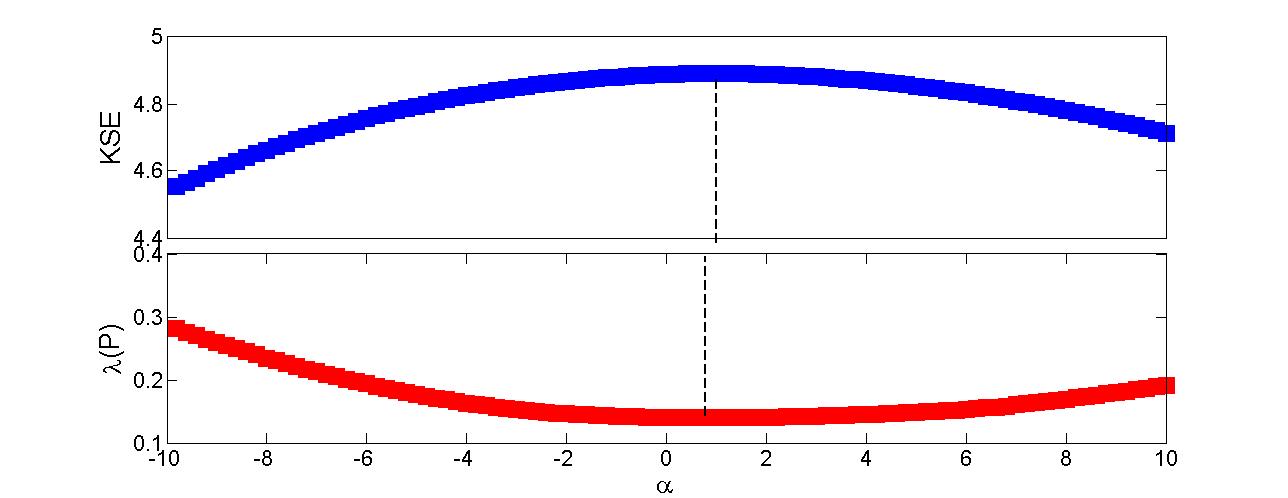}
  \caption{KSE (top) and $\lambda(P)$ (bottom) function of $\alpha$ for a network of size $m=400$ with a proportion of $0$ in $A$ equal to $1/3$. }
\label{fig:KS5}
 \end{figure}

 If $\alpha <0$ we favor transitions towards low degrees nodes, if $\alpha=0$ we find the typical random walk on network  and if  $\alpha>0$ we favor transitions towards high degrees nodes. We assume here that $A$ is symmetric. It may then be checked  that the stationary probability density is equal to:

\begin{equation}
\label{eq:diff2}
 \pi_{stat_i}=\frac{c_ik_i^\alpha}{\sum_j c_jk_j^\alpha},
\end{equation} 
where $c_i=\sum_j A_{ij}k_j^\alpha$,

Using  Eqs.~(\ref{eq:diff1}) and (\ref{eq:diff2}), we check that the transition matrix is reversible and then has $m$ real eigenvalues. From this stationary probability density, we can thus compute both the KSE and the second largest eigenvalue $\lambda(P)$ as a function of $\alpha$. The result is provided in (Fig.~\ref{fig:KS5}).

We observe in (Fig.~\ref{fig:KS5}) that  the KS entropy has a maximum at a value that we denote $\alpha_{KS}$, in agreement with the findings of \cite{gomez2008entropy}. Likewise, $\lambda(P)$ (i.e. the mixing time) presents  a minimum for  $\alpha=\alpha_{mix}$. Moreover, $\alpha_{KS}$ and $\alpha_{mix}$ are close. This means that the two optimal diffusion coefficients are close to each other. Furthermore, looking at the ends of the two curves, we  can find two special Markov chains $P1$  and $P2$ such that $h_{KS}(P1) \leq h_{KS}(P2)$ and $t_1(\epsilon) \leq t_2(\epsilon)$, illustrating that the link between KSE and the minimum mixing time is only true in a general statistical sense.\

We have thus shown that, for a given transition matrix $P$ (or equivalently for given jump rules) the greater the KSE, the smaller the mixing time. We now investigate whether a similar property holds for dynamics, i.e. whether transition rules that maximise KSE are close to the ones minimizing the mixing time. For a given network, i.e. for a fixed $A$, there is a well known procedure to compute the transition matrix $P_{KS}$ which maximizes the KSE with the constraints  $A(i,j)=0 \Rightarrow P_{KS}(i,j)=0$ \cite{burda2009localization}.  It proceeds as follow:  let us note $\lambda$ the greatest eigenvalue of $A$ and $\Psi$  the normalized eigenvector associated  i.e $A\Psi=\lambda \Psi$ and $ \sum_i \Psi^2_i=1$. We define $P_{KS}$ such that:

\begin{equation}
\label{eq:pks}
P_{KS}(i,j)=\frac{A(i,j)}{\lambda}\frac{\Psi_j}{\Psi_i}.
\end{equation}

 We have $ \forall i$ $ \sum_j P_{KS}(i,j) =1$. Moreover, using the fact that $A$ is symmetric we find:

\begin{equation}
\label{eq:sta}
\sum_j P_{KS}(j,i)\Psi^2_j=\sum_j \frac{A(j,i)\Psi_i\Psi_j}{\lambda}=\Psi^2_i.
\end{equation}

 Hence, $P_{KS}^t \Psi^2=\Psi^2$ and the stationary density of $P_{KS}$ is $\pi_{stat}=\Psi^2$.

Using Eqs.~(\ref{eqhks}) and (\ref{eq:pks}), we have:  

\begin{equation}
\label{eq:pks2}
h_{KS}=-\frac{1}{\lambda}\sum_{(i,j)} A(i,j) \Psi_i \Psi_j \log(\frac{A(i,j)}{\lambda}\frac{\Psi_i}{\Psi_j}).
\end{equation}

Eq.~(\ref{eq:pks2}) can be split in two terms:

\begin{eqnarray}
\label{eq:pks3}
h_{KS}&=&\frac{1}{\lambda}\sum_{(i,j)} A(i,j)\Psi_i\Psi_j\log(\lambda)\nonumber\\
&-&\frac{1}{\lambda}\sum_{(i,j)} A(i,j)\Psi_i\Psi_j\log(A(i,j)\frac{\Psi_j}{\Psi_i}).
\end{eqnarray}

The first  term is equal to $\log(\lambda)$ because $\Psi$ is an eigenvector of $A$ and the second term is equal to $0$ due to the symmetry of $A$. Thus:

\begin{equation}
\label{eq:pks4}
h_{KS}=\log(\lambda).
\end{equation}

 Moreover,  for a Markov chain the number of trajectories of length $n$ is equal to  $ N_n=\sum_{(i,j)} (A^n)(i,j)$. For a Markov chain the KSE can be seen as the time derivative of the path entropy leading that KSE is  maximal  when the paths are equiprobable. For an asymptotic long time the maximal KSE is:

\begin{equation}
\label{eq:pks5}
 h_{KS_{max}}=\frac{\log(N_n)}{n} \rightarrow\log(\lambda),
\end{equation}

 by diagonalizing $A$. Using Eqs.~(\ref{eq:pks4}) and (\ref{eq:pks5}) we find that  $P_{KS}$ defined as in Eq.~(\ref{eq:pks}) maximises the KSE.
Finally $P_{KS}$ verifies $\pi_{stat_i}P_{KS}(i,j)=\pi_{stat_j}P_{KS}(j,i)$ $\forall$ $(i,j)$ and thus $P_{KS}$ is reversible.\

In a similar way, we can search for a transition matrix  $P_{mix}$  which minimizes the mixing time -or, equivalently the transition matrix minimizing its second eigenvalue $\lambda(P)$. This problem is much more difficult to solve than the first one, given that the eigenvalues of $P_{mix}$ can be complex. Nevertheless, two cases where the matrix $P_{mix}$ is diagonalizable on $\mathbb{R}$  can be solved \cite{boyd2004fastest}: the case where $P_{mix}$ is symmetric and the case where $P_{mix}$ is reversible for a given fixed stationary distribution.  Let us first consider the case where $P$ is symmetric. The minimisation problem takes the following form:

\begin{equation}
\label{eq:mixmix}
\left\{
  \begin{array}{rcr}
   \min\limits_{P \in S_n} \lambda(P) \\
    P(i,j) \geq 0, P*\textbf{1}=\textbf{1} \\
  A(i,j)=0 \Rightarrow P(i,j)=0\\
  \end{array}
\right.
\end{equation}

 given the strict convexity of $\lambda$ and the compactness of the stochastic matrices, this problem admits an unique solution.

 $P$ is symmetric thus $\textbf{1}$  is an eigenvector associated with the largest eigenvalue of $P$. Then the eigenvectors associated to  $\lambda(P)$ are in the orthogonal of $ \textbf{1}$.The orthogonal projection on $\textbf{1}^{\perp}$ writes:
 $ I_d-\frac{1}{n}\textbf{1}\textbf{1}^t$

%De plus si l'on prend la norme matricielle associée à la norme euclidienne c'est à dire pour $M$ matrice quelconque $|||M|||= \max \frac{ ||MX||_2}{||X||_2} \text{ } X \in \mathbb{R}^n \text{ } X\neq 0$ alors celle-ci est égale à la racine carrée de la plus grande valeur propre de
% $ MM^t$ et donc si $M$  est symétrique elle vaut $\mu(M)$.

Moreover, if we take the matrix norm associated with the euclidiean norm i.e.  for $M$ any matrix $|||M|||= \max \frac{ ||MX||_2}{||X||_2} \text{ } X \in \mathbb{R}^n \text{ } X\neq 0$   it is equal to the square root of the largest eigenvalue of  $ MM^t$ and then if $M$  is symmetric it is equal to $\lambda(M)$.

Then the minimization problem can be rewritten:

\begin{equation}
\label{eq:mix2}
\left\{
  \begin{array}{rcr}
   \min\limits_{P \in S_n} ||| (I_d-\frac{1}{n}\textbf{1}\textbf{1}^t)P(I_d-\frac{1}{n}\textbf{1}\textbf{1}^t)|||=|||P-\frac{1}{n}\textbf{1}\textbf{1}^t|||\\
    P(i,j) \geq 0, P*\textbf{1}=\textbf{1} \\
  A(i,j)=0 \Rightarrow P(i,j)=0\\
  \end{array}
\right.
\end{equation}

We solve this constrained optimization problem (Karush-Kuhn-Tucker) with Matlab and we denote $P_{mix}$ the matrix which minimizes this system.\

 We remark that the mixing time of $P_{KS}$ is smaller than the mixing time of $P_{mix}$. This is coherent because in order to calculate $P_{KS}$ we take the minimum on all the matrix space whereas to calculate $P_{mix}$ we restrict us to the symmetric matrix space. Nevertheless, we can go a step further and calculate, the stationary distribution being fixed, the reversible matrix which minimizes the mixing time.  If we note $\pi$ the stationary measure and $\Pi=diag(\pi)$. Then $P$ is reversible if and only if $\Pi P=\Pi^t P$. Then in particular $\Pi^{\frac{1}{2}}P\Pi^{-\frac{1}{2}}$ is symmetric and has the same eigenvalues as  $\Pi$. Finally, $p=(\sqrt{\pi_1},...,\sqrt{\pi_n})$ is an eigenvector of $\Pi^{\frac{1}{2}}P\Pi^{-\frac{1}{2}}$ associated to the eigenvalue $1$. Then the minimization problem can be written as the following system:

\begin{equation}
\label{eq:mix2}
\left\{
  \begin{array}{rcr}
   \min\limits_{P} ||| (I_d-\frac{1}{n}\textbf{q}\textbf{q}^t)\Pi^{\frac{1}{2}}P\Pi^{-\frac{1}{2}}(I_d-\frac{1}{n}\textbf{q}\textbf{q}^t)|||\\
   =|||\Pi^{\frac{1}{2}}P\Pi^{-\frac{1}{2}}-\frac{1}{n}\textbf{q}\textbf{q}^t|||\\
    P(i,j) \geq 0, P*\textbf{1}=\textbf{1}, \Pi P=\Pi^t P  \\
  A(i,j)=0 \Rightarrow P(i,j)=0\\
  \end{array}
\right.
\end{equation}

\begin{figure}[hb!]
   \centering
   \includegraphics[width=10cm]{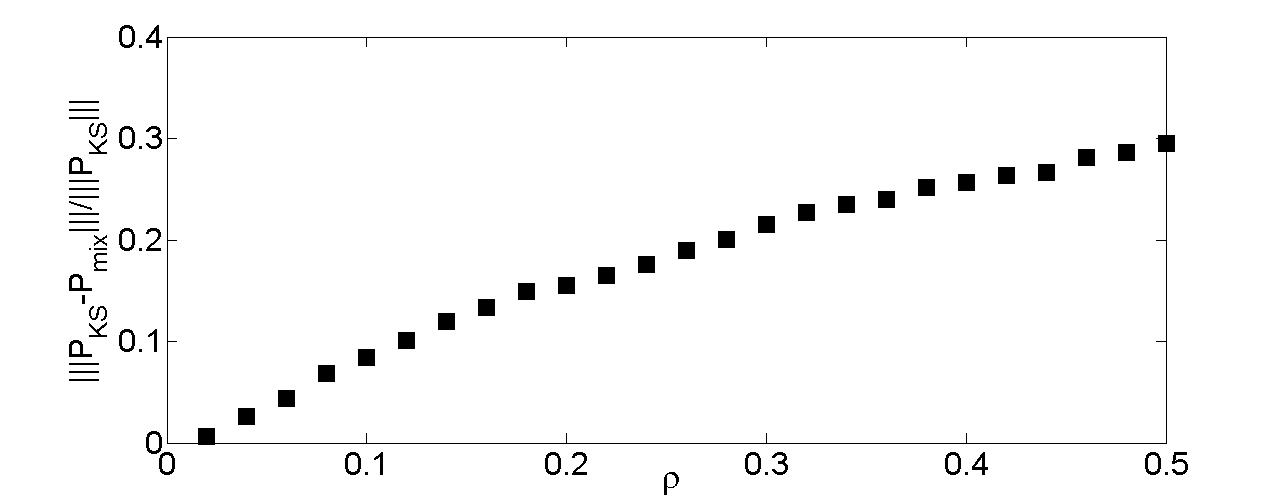}
  \caption{$|||P_{KS}-P_{mix}|||/|||P_{KS}|||$ as a function of the density $\rho$ of $0$ present in $A$.}
\label{fig:KS2bis}
 \end{figure}

When we implement this problem in Matlab with $\pi=\pi_{KS}$ we find a  matrix $P_{mix}$  such that naturally $\lambda(P_{mix}) \leq  \lambda(P_{KS})$. Moreover we can compare both dynamics by evaluating $|||P_{KS}-P_{mix}|||$ compared to $|||P_{KS}|||$ which is approximatively equal to $|||P_{mix}|||$. We remark that $|||P_{KS}-P_{mix}|||$ depends on the density $\rho$ of $0$ in the matrix $A$. For a density equal to $0$ the matrices $P_{KS}$ and $P_{mix}$ are equal and the quantity $|||P_{KS}-P_{mix}|||$ will increase continuously when $\rho$ increases. This is shown in (Fig.~\ref{fig:KS2bis}).

From this, we conclude that the rules which maximize the KSE are close to those which minimize the mixing time. This becomes increasingly accurate as  the fraction of removed links in $A$ is weaker. Since the calculation of $P_{mix}$ quickly becomes tedious for quite large values of $m$, we offer here a  much cheaper alternative by computing  $P_{KS}$ instead of $P_{mix}$. \

Moreover, maximizing the KSE appears today as a method to describe out of equilibrium complex systems \cite{monthus2011non},  to find natural behaviors \cite{burda2009localization} or to define optimal diffusion coefficients in diffusion networks. 
This general observation however provides a possible rationale for selection of stationary states in out-of-equilibrium physics: it seems reasonable that in a physical system with two simultaneous equiprobable possible dynamics, the final stationary state will be closer to  the stationary state corresponding to the fastest dynamics (smallest mixing time). Through the link found in this letter, this state will correspond to a state of maximal KSE. If this is true, this would provide a more satisfying rule for selecting stationary states in complex systems such as climate than the maximization of the entropy production, as already suggested in \cite{mihelich2014maximum}.

{\bf Acknowledgments} Martin Mihelich thanks IDEEX Paris-Saclay for financial support. Quentin Kral was supported by the French National Research Agency (ANR) through contract ANR-2010 BLAN-0505-01 (EXOZODI).\

%\nocite{*}
 \bibliographystyle{unsrt}
 \bibliography{biblio}

\end{document}